\documentclass[12pt]{elsarticle}
\usepackage[utf8]{inputenc}
\usepackage{amssymb}
\usepackage{graphicx} 
\graphicspath{{./Figuras/}}
\usepackage[hidelinks]{hyperref}
\usepackage{natbib}

\journal{Arxiv}

\begin{document}
\begin{frontmatter}


\title{Could spectral converters really enhance photovoltaics?}


\author{Marcos Paulo Belan\c{c}on$^a$} 
\ead{marcosbelancon@utfpr.edu.br}

 \address[$^a$]{Pós-graduação em processos químicos e bioquímicos, Universidade Tecnol\'{o}gica Federal do Paran\'{a}, Via do Conhecimento Km 1, 85503-390 Pato Branco - Brazil}

\begin{abstract}
Spectral converters\cite{Ahmed2017,Fetlinski2018,Grigoroscuta2018} have been proposed as a tool to achieve high efficiencies from Silicon photovoltaics(PV). Even though from the scientifical point of view it could be interesting to acomplish that, scientists should concentrate effort to invest in materials that can provide more return: in the case of spectral converters, they are very unlikely to have a place in flat-panel PV's. However, concentrated photovoltaics (CPV) technology may get a boost from such devices, an more attention should be driven to this goal.
\end{abstract}


\end{frontmatter}


\section{Introduction}
\label{sec:intro}

Single junction Silicon solar cells share $>90$\% of the world photovoltaic production\cite{ITRPV2015}. Scientists have been trying to develop newer technologies into the industry level, such as thin-films\cite{Fthenakis2012}. By improving energy conversion, cost and scale of production\cite{Summary2016} they aim by one hand to reduce land use and energy cost. By another hand, there is a growing worry about the future availability of key raw materials that could constraint the maximum production scale of some technologies\cite{Feltrin2007,Fthenakis2009,Graedel2011}. Even Silicon PV's may face some constraint, mainly due Silver\cite{TheSilverInstitute2014}, which is the most common material choice of industry to build the thin wires that collect the current from the surface of PV cells.

The last decade of subsidies speed up the pass of industry innovation, and every single piece of the Silicon PV supply chain have been optimized to reduce cost\cite{ITRPV2015}. Since 2011, the price of polycristalline Silicon has dropped by 75\%; average polysilicon per wafer (156x156$mm^2$) is expected to decrease 20\% in the next decade mainly by reducing cell thickness, as well Silver consumption is expected to be cutted by half. Even front glasses in modules are being optimized to be thinner, all of that reducing materials consumption, weight, etc, pushing prices down. These efforts from industry have created a huge challenge for scientists. The technologies being developed should be more efficient, while avoiding scarse minerals, be able to reach Terawatt scale of production and compete with Silicon PV's that have experienced a production learning curve as long as few decades.

\section{Economics of spectral converters}

One approach oftenly discussed in the literature is to develop spectral converters\cite{Huang2013c,McKenna2017,DelaMora2017a}, mainly to be used in Silicon PV`s. Theoretically, between the ultraviolet and the near infrared, due the mismatch between Silicon bandgap and the solar spectrum, there is 150 $W/m^2$ of sunlight that could be ``downconverted'' and 160 $W/m^2$ that could be ``upconverted''\cite{DelaMora2017a}. A few different mechanisms could be explored to enhance ``energy density'' in Silicon PV's, the most obvious one should be by ``quantum cutting'' one UV/Blue photon in two infrared photons, where it is more likely to have photons converted into electric current in the Silicon PV.

Even though promising, such approaches have to be carefully criticized. Silicon PV's today can harvest about 450$W/m^2$, which is translated in the fact that Silicon solar cell efficiency is limited at about 30\%, and a downconverting layer could improve this number to not more than 38.6\%\cite{Richards2006}. From the industry point of view, such layer could cost at maximum 1/3 of the PV module price; i.e. Silicon PV at the industry scale level can already produce 300$W/m^2$\cite{ITRPV2015}, which could be improved by an ideal downconverting layer to 400$W/m^2$. If we consider the manufacturing module cost of about 0.31 US\$/$W_p$ experienced by industry last year, one could conclude that a downconverting layer can make sense at the industry level only if its costs can go below 30 US\$ per module, or about 20 US\$/$m^2$.

Each mm of glass in the front surface of the module weights about 2.5kg/$m^2$, what means that if we aim to employ scarse, high purity and expensive materials to produce a downconverting layer, it should be a thin-film. By this way, a layer one micrometer thick, would consume 2-3 grams of raw material per $m^2$ if its density is similar to that of borosilicate glass. So, at around 10 US\$ a gram, what are the raw materials available for us to help industry push Silicon PV towards its limits?

\section{Photovoltaics versus Concentrated photovoltaics in the future}

In the literature, a few different thin films have been investigated to work as downconverting layer in Silicon PV. For example, Elleuch et al\cite{Elleuch2015} have succesfully prepared downconverting antireflective thin films based on ZnO:$Er^{3+},Yb^{3+}$, and Lin et al\cite{Lin2017a} demonstrated an impressive small reflectivity of only 1.46\% in $Y_2O_3$:$Bi^{3+},Yb^{3+}$ thin films. In both works the improvement of Silicon PV efficiency was not measured, even though the $Yb^{3+}$ emission around 980 nm have been proposed as a path to achieve downconversion of UV/VIS photons. In any case, high purity raw materials are needed and complex techniques are employed to produce the layers. To reach TW scale of production in the future, we cannot rely on such approachs if we take into account that at industry level we can spend just a few dollars per square meter. 

Commercial antireflective (AR) coatings today have a lifetime of about 10 years\cite{ITRPV2015}, what is expected to double in the next decade; and they are not spectral converters. So, even if we could make these AR-coatings become spectral converters without introducing any cost, the short lifetime would make the final energy production of the panel just marginally higher.

By the considerations presented here, the conclusion we have found could not be another: there is no room for spectral converters in Silicon PV's at the industry level. Scientists working with materials for solar energy should pay attention into that and drive the efforts in looking for breakthroughs instead of very unlikely marginal enhancement of PV's efficiencies. 

Concentrated photovoltaics (CPV) have not been widely employed today, the projects are more expensive than conventional flat-panels. However, in CPV's light can already be concentrated 1000 times\cite{Philipps2015}, a factor that could be even higher from the physical point of view. That is what opened the room for III-V multi-junction Solar Cells to reach the market, and may be spectral converters could be a smart option to develop such projects.

Instead of expend at maximum a few dollars per square meter of layer in a conventional PV, industry is able to spend a few thousand dollars  in a layers for CPV's; by this way, expensive raw materials and process could be deployed. Additionally, one may point out that many technologies today are already facing constraints due mineral scarcity (Te, Ge, Ga, In, Ag, etc)\cite{Philipps2015,Kannan2016,Lee2017,Graedel2011,Summary2016,Grandell2016}, and CPV's could achieve a much better ratio between \textit{energy produced and minerals consumed}.

\section{Acknowledgments}
\label{S:5}
The authors would like to thank Brazilian agency CNPq (grant $480576/2013-0$) and CAPES for their financial support.





\bibliographystyle{model1-num-names}
\bibliography{/home/mbelancon/Documentos/library}







\end{document}